\documentclass[twocolumn,showpacs,superscriptaddress,preprintnumbers,amsmath,amssymb]{revtex4-2}
\usepackage{graphicx}
\usepackage{dcolumn}
\usepackage{appendix}
\usepackage{bm}

\usepackage[usenames,dvipsnames]{xcolor}
\usepackage{subfigure}
\usepackage{bbm}
\usepackage{enumerate}
\usepackage[
  bookmarks=true,
  colorlinks,
  linkcolor=blue,
  urlcolor=blue,
  citecolor=blue,
  plainpages=false,
  pdfpagelabels,
  final,
  breaklinks=true
]{hyperref}

\usepackage{titlesec}
\usepackage{array}
\usepackage{dsfont}
\usepackage{physics}
\usepackage{mathpazo}

\begin{document}

\title{Entanglement certification of attosecond light fields with photoelectron interferometry}

\author{Philipp Stammer}
\email{philipp.stammer@icfo.eu}
\affiliation{ICFO -- Institut de Ciencies Fotoniques, The Barcelona Institute of Science and Technology, 08860 Castelldefels (Barcelona), Spain}
\affiliation{Atominstitut, Technische Universit\"{a}t Wien, 1020 Vienna, Austria}

\date{\today}

\begin{abstract}

The emergence of entangled quantum optical states in high-harmonic generation raises a fundamental problem about its characterization, namely, \textit{how can entanglement of extreme-ultraviolet radiation be certified for attosecond light fields?} Here, we provide experimentally accessible separability tests, particularly showing that attosecond photoelectron interferometry provides a direct solution to this problem. This is achieved by a quantum optical treatment of mapping photoelectron interferometry to the measurement of optical correlation functions. This yields a negative-partial-transpose witness for infrared-XUV entanglement and a sufficient witness of genuine multipartite entanglement without auxiliary optical measurements. We evaluate both witnesses on explicit optical states and identify how the protocol can be extended to XUV-XUV entanglement certification. Hence, we establish attosecond photoelectron interferometry as a general framework for entanglement certification across a wide range of optical frequencies.

\end{abstract}

\maketitle

\section{Introduction}

Entanglement between optical field modes is emerging as a new fundamental insight in attosecond and strong-field quantum optics, connecting quantum state engineering with radiation spanning from the infrared (IR) to the extreme ultraviolet (XUV) regime~\cite{stammer2025colloquium, cruz2024quantum}. The upconversion process of high harmonic generation (HHG) provides a natural setting for creating such correlations across widely separated frequencies~\cite{stammer2024entanglement, yi2024generation}, however, their experimental identification requires schemes that go beyond conventional approaches. This raises a central question: \emph{how can entanglement of the attosecond optical field itself be certified using experimentally accessible observables?}

Conventional optical entanglement certification schemes combine measurements of field quadratures or photon correlations with separability criteria, either through quantum state reconstruction or by witnesses involving carefully selected observables~\cite{Lvovsky2009, GuhneToth2009, friis2019entanglement, HilleryZubairy2006, van2003detecting}. 
For Gaussian states, first and second quadrature moments completely specify the state, enabling entanglement analysis through covariance matrices~\cite{weedbrook2012gaussian, serafini2023quantum}. For non-Gaussian states, however, entanglement can escape detection at the covariance level. Moment-based approaches address this limitation by imposing separability constraints on expectation values of products of creation and annihilation operators, including higher-order field coherences and photon-number correlations~\cite{HilleryZubairy2006, shchukin2005inseparability}. Violating such inequalities can certify entanglement without full tomography. Nevertheless, their practical usefulness depends on the experimental accessibility of the required moments, and if a measurable subset can distinguish entangled states from separable alternatives.
A state is entangled across a specific bipartition when it cannot be expressed as a statistical mixture of product states across that partition~\cite{GuhneToth2009, bertlmann2023modern}. For three or more modes, genuine multipartite entanglement (GME) requires excluding all biseparable mixtures, including those whose components factorize across different bipartitions~\cite{GuhneSeevinck2010, huber2010detection, baksova2025multi}. 
Consequently, for mixed states, demonstrating entanglement across every fixed bipartition does not by itself establish GME~\cite{teh2014criteria, leskovjanova2026existence}. This distinction is essential for identifying quantum correlations shared among several optical modes, and therefore continuous-variable GME criteria require bounds valid for the entire convex set of biseparable states~\cite{teh2014criteria}.
For broadband quantum light, entanglement certification is particularly challenging due to the missing experimentally accessible field correlations for constructing such bounds.

\begin{figure}
    \centering
	\includegraphics[width = 1.0 \columnwidth]{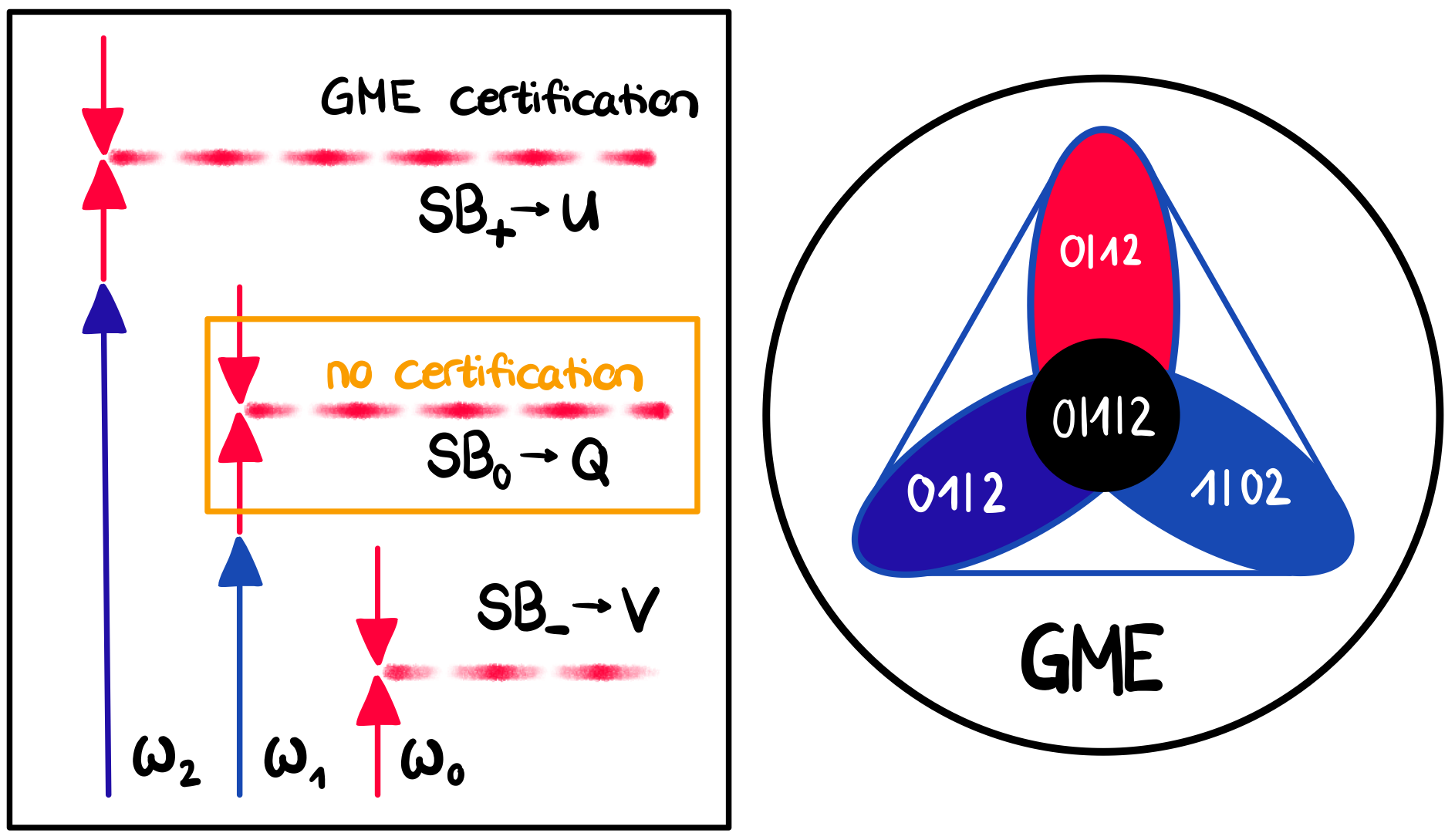}
	\caption{\textbf{Entanglement certification of light using quantum optical attosecond photoelectron interferometry.} Left: Quantum optical RABBIT gives rise to field correlation functions $(Q,U,V)$ from the measurement of photoelectron sideband ($\text{SB}$). While a single sideband is insufficient to certify entanglement, the combination of the central sideband $\text{SB}_0$ with an upper and lower band $\text{SB}_\pm$ provides access to various entanglement certification bounds. 
    Right: Schematic separability structure for the field modes $(012)$. Genuine tripartite entanglement certification excludes the convex hull of the three sets separable across the bipartitions $(0|12)$, $(1|02)$, and $(01|2)$, including mixtures of different cuts. 
    }
    \label{fig:cartoon}
\end{figure}

In strong-field quantum optics~\cite{stammer2023quantum, lewenstein2021generation, thesis_stammer}, these questions arise because the exchange of photons between the field modes is correlated~\cite{stammer2024information}. Quantum-optical descriptions predict squeezing and entanglement between these modes~\cite{stammer2022high, stammer2024entanglement, yi2024generation, gombkotHo2026intermodal, stammer2022theory, rivera2024role, de2025second, yan2026entanglement, stammer2026high, rivera2026emergence}, while experimental signatures of non-classical correlations between low-order harmonics have been reported~\cite{theidel2024evidence}. Beyond the fundamental significance of generating entangled quantum states in HHG, correlations between photon frequencies offer additional control over spectroscopic signals of matter~\cite{Dorfman2016, Rahav2010}. 
Therefore, certification is particularly important for distinguishing entanglement from other field correlations.
However, transferring conventional optical characterization techniques to the XUV regime remains experimentally challenging, particularly for measurements requiring controlled mode-matched local oscillators for phase-sensitive coherences or efficient detection of photon-number correlations across widely separated frequencies. However, attosecond spectroscopy offers an alternative route through photoelectron interferometry~\cite{veniard1996phase}. Reconstruction of attosecond beating by interference of two-photon transitions (RABBIT) converts the interference between different IR-XUV ionization pathways into a delay-dependent electron signal~\cite{muller2002reconstruction}, see Fig.~\ref{fig:cartoon}, and have been used for the measurement of classical attosecond pulse properties~\cite{Paul2001}. Now, recent quantum-optical treatments show that its sideband amplitudes and phases encode field correlation functions, including multimode coherences~\cite{dubois2026quantum}. This connection makes photoelectron spectra sensitive to quantum light, but an observed interference fringe does not by itself allow for entanglement certification -- simply because coherent product states interfere as well. Instead, entanglement certification requires more, and proven certification bounds do yet not exist for attosecond light fields.

Here we establish such bounds for RABBIT and identify the necessary electron observables to evaluate them. We show why an ordinary single-sideband fringe is insufficient for entanglement certification, but continue to construct bounds that can certify IR-XUV entanglement and GME among two XUV modes and the IR field. We show that the key ingredient is using multiple sideband channels that allow to measure the required moments from electron detection alone without auxiliary optical measurements. 
These results establish attosecond photoelectron interferometry as an experimentally accessible route to certify optical entanglement.

\section{The problem of entanglement certification of light using photoelectrons}

The principle underlying the entanglement certification scheme is based on the measurement of photoelectrons after ionization from an XUV field of frequency $\omega_i$, with $i = 1,2$ (see Fig.~\ref{fig:cartoon}), and photon exchange with an IR field at time-delay $\tau$. In schemes such as RABBIT the resonant condition is given by $\omega_2  = \omega_1 + 2 \omega_0$, with the IR frequency $\omega_0$.
The transition operators corresponding to a given RABBIT sideband are described by 
\begin{align}
    T_{0} (\tau) = g_1 a_1 a_0 e^{i \omega_0 \tau } + g_2 a_2 a_0^\dagger e^{- i \omega_0 \tau},
\end{align}
where the bosonic field operators $a_i$ obey canonical commutation relation $[a_i,a_j^\dagger] = \delta_{ij}$ for $i,j = 0,1,2$. 
Consequently, the electron yield of the RABBIT sideband (SB) is obtained via $\operatorname{SB}_0(\tau) \equiv \expval*{T_{0}^\dagger T_{0}}$, and given by
\begin{align}
\label{eq:SB0}
    \operatorname{SB}_0(\tau)  = g_1^2D_1 + g_2^2D_2 +  g_1 g_22 \operatorname{Re}\left[ Qe^{2 i \omega_0 \tau} \right],
\end{align}
where the field correlations, including the photonic number operators $N_i=a_i^\dagger a_i$, are encoded in 
\begin{align}
    D_1 =  \expval{N_1 N_0}, \,
    D_2 =  \expval{N_2(N_0 + 1)}, \, 
    Q  =  \expval*{a_2^\dagger a_1 a_0^2}. 
\end{align}

These quantities encode joint three-mode optical coherence properties of the two XUV modes and the IR field, where the oscillating coherence from the sideband encode the correlations of $Q$. Note that the channel involving the IR emission has the bosonic $N_0+1$ factor due to the additional vacuum contribution. 

Although the RABBIT fringes encode information about the field correlations, the signal is insufficient as an entanglement witness when used at a single sideband. This can be seen when defining the operators $A=a_2 a_0^\dagger$ and $B=a_1 a_0$, and since $Q = \expval*{A^\dagger B}$, it follows from the Cauchy-Schwarz inequality that $\abs{Q}^2 \leq D_1 D_2$. This holds for every optical state, entangled or separable~\cite{dubois2026quantum}. An even stronger statement is that the set of observables $\{ Q, D_1, D_2 \}$ admits a fully separable explanation for every tuple possible, given by a squeezed vacuum IR with $\expval*{N_0}=1$ and $\expval*{a_0^2}=\sqrt{2}$ plus a coherent upper harmonic and weighted mixture of opposite phases lower harmonic with mean values $\expval*{N_2} = D_2/2$ and $\expval*{N_1} = D_1$, respectively.
This construction establishes the statement that \emph{no criterion based only on the set of observables from a single sideband can certify entanglement without further information}.
However, RABBIT does more than providing a single sideband, and in the following we shall see how multiple sidebands provide the important correlation measurements. We will see that the path interference provides the necessary coherence measurements, while the derived certificate provides the separability threshold.

\section{Entanglement certification with RABBIT}

The central sideband $\operatorname{SB}_0$ given by Eq.~\eqref{eq:SB0} is obtained from the interference of the harmonic $\omega_1$ plus absorption of a single IR photon and the harmonic $\omega_2$ plus the emission of an IR photon. 
However, in a measured RABBIT spectrogram we also have the sidebands $\operatorname{SB}_{\pm}$, which are obtained from the transition operators
\begin{align}
    T_{+}(\tau) & = g_2 a_2 a_0 e^{i \omega_0 \tau } + g_{+} a_+ a_0^\dagger e^{ - i \omega_0 \tau}, \\
    T_{-} (\tau ) & = g_- a_- a_0 e^{i \omega_0 \tau} + g_1 a_1 a_0^\dagger e^{- i \omega_0 \tau}.
\end{align}

These additional sidebands allow to obtain the quantities (see Appendix~\ref{app:3sideband})
\begin{align}
    U  = \expval*{N_2N_0}, \quad V  = \expval*{N_1(N_0 + 1)}.
\end{align}

With this, we obtain the XUV-IR entanglment certification bound 
\begin{align}
\label{eq:bound_IR-XUV}
    \abs{Q}^2 > UV, 
\end{align}
which can be proven with the negative partial transpose.

\emph{Proof.}
For a PPT state let $\sigma = \rho^{T}$ be the partial transpose with respect to the IR field mode, with every separable state $\sigma \geq 0$. Choose $f_1 = a_2 a_0$ and $f_2 = a_1a_0^\dagger$. Positivity of $\sigma$ implies that the moment matrix $G_{ij} = \operatorname{Tr}[\sigma f_i^\dagger f_j]$ is positive semi-definite.  
Transposition in the IR number basis gives $N_0^T = N_0$ and $[(a_0^\dagger)^2]^T = a_0^2$, and consequently
\begin{align}
    G = 
    \begin{pmatrix}
    \expval*{N_2 N_0} & Q\\
    Q^* & \expval*{N_1(N_0+1)}
    \end{pmatrix} =
    \begin{pmatrix}
        U  & Q \\
        Q^* & V 
    \end{pmatrix} \succeq 0 .
\end{align}

Its determinant must satisfy
\begin{align}
    \operatorname{det}G = UV - \abs{Q}^2 \geq 0.
\end{align}

Violation therefore certifies a negative partial transpose, and therefore NPT entanglement across the mode decomposition $0(12)$. This implies entanglement between the IR and the XUV modes.  
Note that the number-operator ordering is dictated by the proof, and hence it cannot be replaced by the measurable $D_1$ and $D_2$, whose product gives only the universal bound that does not allow certification.  
Importantly, this does not contradict the no-go result from the tuple $\{Q, D_1,D_2 \}$, because these quantities are obtained from the fixed central sideband. However, the outer sidebands provide the additional moments $U,V$. Consequently, the full three-sideband dataset contains the necessary information to construct the entanglement certification bound~\cite{shchukin2005inseparability}.
This certificate excludes separability between the IR and the joint harmonic subsystem. It does yet not exclude a state in which one harmonic factorizes from an entangled IR-harmonic pair. Genuine tripartite certification must additionally exclude mixtures separable across any of the three cuts.

\section{Genuine multipartite entanglement certification}

So far we have derived a certification bound for the entanglement between the IR and the joint XUV field. However, the optical field consists of three modes, and hence, the natural extension is to look for genuine multipartite entanglement (GME) bounds. In the following we will provide a criterion that excludes arbitrary mixtures that are separable across different cuts. 
Consider the modes $(012)$, and the available measurement of $\{ Q, U, V\}$ from the three-sideband spectrogram. 
Given access to these quantities, for every biseparable state we have
\begin{align}
\label{eq:GME_bound}
    \abs{Q} & \leq H(U,V) \leq H(U_{dc}, V_{dc}), 
\end{align}
and hence a sufficient GME certification is given by the violation thereof, with the bound given by
\begin{equation}
\begin{aligned}
    H(U,V) = & \sqrt{(2-\sqrt3) V} \\
    & + (1 + \sqrt{2}) \sqrt{U(V+1)} + \sqrt{2 UV}.
\end{aligned}
\end{equation}

Violation of Eq.~\eqref{eq:GME_bound}, and observation of $\abs{Q} > H(U_{dc},V_{dc}) \geq H(U, V)$ certifies GME among the modes $(012)$ from the measurement of the central RABBIT fringe plus the two neighboring sidebands. 
Note that the additional harmonics $a_\pm$ reduce the sensitivity by increasing the certification threshold using $H(U_{dc},V_{dc})$, but nevertheless allow to a certification protocol. The tighter bound given $H(U,V)$ can be achieved by proper spectral shaping of the XUV pulse. 

\emph{Proof sketch.} The idea of the proof is as follows, while the full proof is in the Appendix~\ref{app:GME_proof}. 
The proof essentially identifies that the involved transitions on different Fock sectors $\ket{t,s,r}$ always involve two IR photon transitions. Hence, the contributions to $Q$ are separated into different sectors, such that the coherences can be bounded using positivity of the density matrix and Cauchy-Schwarz inequality. Bounding these contributions for all biseparable states yields the desired bound. 

While the certification bound of Eq.~\eqref{eq:GME_bound} is a sufficient criterion for GME, it is not proven to be optimal. 
Therefore, a particularly interesting case is that of $U=0$, which allows to provide an optimal certification bound from the measured moments alone.
This is because every biseparable three-mode state with $U=0$ satisfies
\begin{align}
\label{eq:GME_bound_optimal}
    \abs{Q}^2 \leq (2 - \sqrt{3}) V,
\end{align}
and hence the bound is optimal for certification of GME of these quantities alone. Hence, for $U=0$,
\begin{align}
    \abs{Q}^2 > (2-\sqrt{3}) V \Longleftrightarrow (Q,V) \text{ excludes biseparability}.
\end{align}

This result is particularly interesting when the upper sideband probing $U=\expval*{N_2 N_0}$ exclude joint IR and XUV population, such that the contributions to the coherence $Q$ are exclusively from transitions between states of the form $\ket{2,j,0} \leftrightarrow \ket{0, j-1,1}$, $j \geq 1$. This transition also occurs in biseparable states such that $Q \neq 0$ alone does not certify GME, but in contrast is bounded by Eq.~\eqref{eq:GME_bound_optimal}. 
Therefore, at $U=0$, the criterion is optimal for the measured moments, and its violation is necessary and sufficient to exclude every biseparable explanation of $(Q,V)$. 
An explicit construction of such a biseparable state reproducing every point inside the allowed disk including its boundary is given by $\ket{\alpha}_1 \otimes [\sqrt{1-z_*}\ket{0,1}_{02} + e^{i \phi}\sqrt{z_*} \ket{2,0}_{02}]$ for $z_* = (\sqrt{3} -1)/2$. Choosing $\abs{\alpha}^2 = V/(1+2z_*)$ gives $Q =e^{i\phi} \sqrt{(2-\sqrt{3})V}$. Now mixing the opposite phases, $\phi$ and $\phi+\pi$, fills the whole disk $\abs{Q}^2 \leq(2-\sqrt{3})V $. This establishes exact optimality for the tuple $(Q,V,U=0)$.

\section{Certifying attosecond light entanglement}

We have seen that attosecond photoelectron interferometry measurements, such as RABBIT allow to construct entanglement certification bounds. 
Motivated by the entangled states from HHG, we show different classes of entangled field states that the newly established bounds are able to detect. However, exploration of the entire space of states that can be certified from these bounds is beyond the scope of the current work. 
First, we show the certification of the IR-XUV entanglement, with the bound established in Eq.~\eqref{eq:bound_IR-XUV}, and consider a displaced two-mode squeezed vacuum field~\cite{loudon1987squeezed}. Such a field is predicted in HHG, where the IR mode is correlated with the harmonic
\begin{align}
\label{eq:XUV-IR_example}
    \ket*{\Phi_{01|2}} = D_0(\alpha) \sqrt{1-\abs{z}^2} \sum_{k=0}^\infty z^k \ket{k,k}_{01} \otimes \ket{\beta}_2,
\end{align}
where $D_0(\alpha)$ is the IR displacement and $z = \tanh (r)$ the squeezing parameter with $r \geq 0$. For the second harmonic mode we consider a classical coherent state. 
When we define $x = \abs{\alpha}^2$ and $n= \abs{z}^2/(1-\abs{z}^2)$, we find that $\abs{Q}^2 = 4 \abs{\beta}^2 x n (n+1)$, $U= \abs{\beta}^2(x+n)$ and $V = n(x+2n+2)$, such that the states certified by RABBIT require $0<n<(4\sqrt{2}-2)/7$ and lie in the interval $x_- < x < x_+$ with
\begin{align}
    x_\pm (n) = \frac{n+2\pm \sqrt{4-4n -7n^2}}{2}.
\end{align}

The Gaussian character of Eq.~\eqref{eq:XUV-IR_example} provides an independent benchmark for the sensitivity of the RABBIT witness since Eq.~\eqref{eq:bound_IR-XUV} tests positivity of a selected moment matrix of the partially transposed state, and a nonnegative determinant alone does not establish separability. For the Gaussian two-mode subsystem $(01)$, in contrast, the covariance-matrix PPT criterion is necessary and sufficient for separability~\cite{simon2000peres, werner2001bound}. The independent upper harmonic does not alter entanglement across $(01|2)$, and the local IR displacement changes only first moments. Using quadratures with vacuum variance $1/2$, the smallest symplectic eigenvalue after partial transposition is
\begin{equation}
    \widetilde{\nu}_- =\frac12e^{-2r} =\frac12\left(\sqrt{n+1}-\sqrt n\right)^2.
\end{equation}

\begin{figure}
    \centering
	\includegraphics[width = 1.0 \columnwidth]{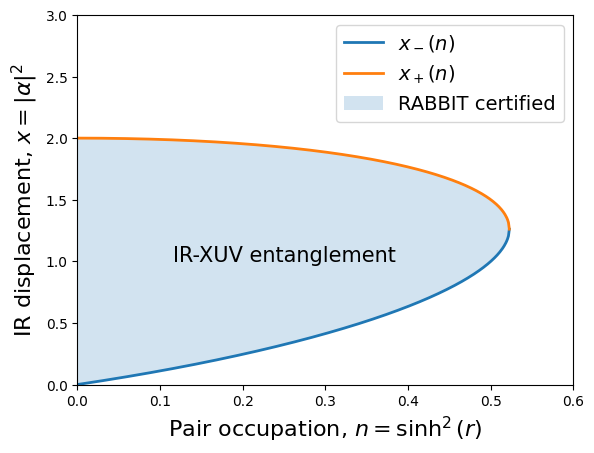}
	\caption{\textbf{Certification of IR-XUV entanglement.} 
    We consider the state shown in Eq.~\eqref{eq:XUV-IR_example}, given by a displaced two-mode squeezed state across the $(01)$ modes, and an independent upper XUV mode $(2)$. The shaded region satisfies $|Q|^2 > UV$, and certifies IR-XUV entanglement from the RABBIT moments.
    It is bounded by $x_-(n) < x < x_+(n)$ for $0 < n < (4\sqrt2-2)/7$, where $n=\sinh^2r$ is the pair occupation and $x=|\alpha|^2$ is the IR displacement intensity. The boundaries correspond to equality and are not certified. The exact Gaussian PPT test identifies entanglement across $(01)$ for every $n>0$ and every $x$.
    }
      \label{fig:IR-XUV}
\end{figure}

Since $\widetilde{\nu}_-<1/2$, it detects NPT entanglement for every $n>0$, independently of $x$. Fig.~\ref{fig:IR-XUV} therefore compares the region certified by the accessible RABBIT moments within the larger region in which the state is entangled.

\begin{figure}
    \centering
	\includegraphics[width = 1.0 \columnwidth]{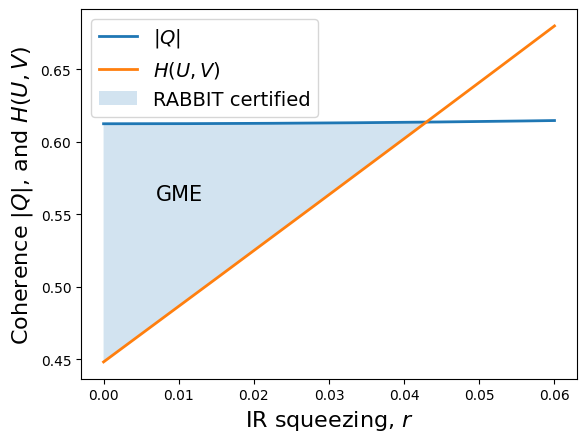}
	\caption{\textbf{Genuine multipartite entanglement certification.} 
    GME certification for the non-Gaussian family of states in Eq.~\eqref{eq:GME_example} at $p=0.75$ and finite moment $U$.
    Shading marks GME certified region $|Q|>H(U,V)$, occurring for $0 \le r < r_c$ with $r_c \simeq 0.04294$. The state remains genuinely tripartite entangled for all $r$ since the squeezing is local.
    }
      \label{fig:GME}
\end{figure}

Moreover, Eq.~\eqref{eq:XUV-IR_example} factorizes across $(01|2)$, and is therefore not GME. To test the genuine tripartite certification of Eq.~\eqref{eq:GME_bound}, we now consider the following non-Gaussian family of states
\begin{align}
\label{eq:GME_example}
    \ket{\Phi_{\text{GME}}} = S_0(r) \left[ \sqrt{p}\ket{0,0,1} + \sqrt{1-p} \ket{2,1,0} \right],
\end{align}
where $S_0(r)$ is IR squeezing operator with $n = \sinh^2(r)$. Given the RABBIT moment based criteria from Eq.~\eqref{eq:GME_bound} we find $\abs{Q} = \sqrt{2p(1-p)}(1+n)$, $U=pn$ and $V=(1-p)(3+5n)$, which are used to evaluate $H(U,V)$. In fact, the state is GME for the entire $0<p<1$, and we show the GME certification in Fig.~\ref{fig:GME} for $p=0.75$ and varying IR squeezing strength $r$.
The shaded region is the $\abs{Q} > H(U,V)$ and indicates where GME can be certified from the RABBIT moment measurement. Note that the state in Eq.~\eqref{eq:GME_example} is not the only set of states that can be detected with the RABBIT criterion, and further exploration of the family of detectable GME states is crucial.

\section{Discussion}

In conclusion, we have developed an attosecond photoelectron interferometry approach that allows for the certification of optical entanglement. In particular, entanglement between the IR and XUV field modes as well as genuine multipartite entanglement between three modes. The certification bounds were tested for two illustrative state families, each violating the respective bound, showing the ability to certify entanglement of attosecond quantum light using conventional RABBIT measurements on the correlated fields. 
Furthermore, we shall provide an outlook towards entanglement certification between reduced XUV-XUV modes. For such a certification protocol a measurement of the harmonic number correlations $\expval*{N_2N_1}$ are needed, which are not accessible from direct RABBIT spectrograms.  
However, certification of inter-harmonic entanglement is possible with the criteria given by 
\begin{align}
    |\expval*{a_2^\dagger a_1}|^2 > \expval{N_2N_1}, 
\end{align}
where $\expval*{a_2^\dagger a_1}$ can be obtained from standard RABBIT measurement, whereas the harmonic cross-population can be obtained from additional electron coincidence detection (see Appendix~\ref{app:XX_proof}) or auxiliary optical measurements~\cite{HilleryZubairy2006}. 
These results open the way towards certification methods from quantum information theory considered in a new light for strong field quantum optics. This provides a novel perspective to the characterization of broadband quantum light using attosecond measurement techniques~\cite{dubois2026quantum, rivera2026attosecond, stammer2026attosecond}. \\

\begin{acknowledgments}
I gratefully acknowledge discussion with Emilio Pisanty about quantum optical RABBIT during breakfast at a workshop in Dresden in August 2025, and with Anne L'Huillier during my PhD defense in June 2026 on using RABBIT and KRAKEN for quantum optical schemes.
ICFO group acknowledges support from: Ministerio de Ciencia y Innovation Agencia Estatal de Investigaciones (R$\&$D project CEX2019-000910-S, AEI/10.13039/501100011033, Plan National FIDEUA PID2019-106901GB-I00, FPI), Fundació Privada Cellex, Fundació Mir-Puig, and from Generalitat de Catalunya (AGAUR Grant No. 2017 SGR 1341, CERCA program), and MICIIN with funding from European Union NextGenerationEU(PRTR-C17.I1) and by Generalitat de Catalunya and EU Horizon 2020 FET-OPEN OPTOlogic (Grant No 899794) and ERC AdG NOQIA.

\end{acknowledgments}

\bibliography{literatur}{}

\appendix

\section*{Appendix}

\subsection{\label{app:3sideband}Three-sideband witness}

Using the transition operator for the additional sideband $\operatorname{SB}_{\pm}$, the corresponding spectrograms are given by 
\begin{align}
    \operatorname{SB}_{+} & = g_2^2 U + g_+^2 R + g_2g_+ 2 \operatorname{Re} \left[ Q_+ e^{ i 2 \omega_0 \tau} \right],  \\
    \operatorname{SB}_- & = g_1^2 V + g_-^2 L + g_1 g_- 2 \operatorname{Re} \left[ Q_- e^{2 i \omega_0 \tau}\right], 
\end{align}
where we have 
\begin{align}
    U & = \expval{N_2 N_0}, &   V &= \expval{N_1 (N_0 + 1)}, \\
    R & = \expval{N_+(N_0+1)},&   L &= \expval{N_- N_0}, \\
    Q_+ & = \expval*{a_+^\dagger a_2 a_0^2},&   Q_- &= \expval*{a_1^\dagger a_-a_0^2}.
\end{align}

The relevant quantities needed are $U$ and $V$, which can be upper bounded by the constant terms of the respective sidebands
\begin{equation}
\begin{aligned}
    U  \leq U_{dc} := \frac{D_+}{g_2^2}, \quad V  \leq V_{dc} := \frac{D_-}{g_1^2},
\end{aligned}
\end{equation}
where the directly measurable constant terms are $D_+ = g_2^2U + g_+^2 R$ and $D_- = g_1^2 V+ g_-^2L$, which can be extracted from a cycle average of the oscillating sidebands. Crucially, they require no assumption that the neighboring harmonics are independent, coherent or separable.
Given these measurement data, every separable state across the $(0|12)$ partition obeys
\begin{align}
    \abs{Q}^2 \leq UV \leq U_{dc} V_{dc},
\end{align}
and consequently
\begin{align}
    \abs{Q}^2 > U_{dc} V_{dc} ,
\end{align}
certifies IR-XUV entanglement.
However, spectral shaping of the XUV field modes allows to access the direct data $(U,V)$ for the tighter bound $\abs{Q}^2 > UV$ by removing the $(R,L)$ contributions. This is achieved when the two additional harmonics $a_\pm$ are removed such that the sidebands $\text{SB}_\pm$ have a constant term only, corresponding to the case of the absorption (emission) of an IR photon together with the absorption of harmonic $a_2$ ($a_1$).
Importantly, the transitions are of the same order as the central sideband and given by
\begin{align}
    T_+' = g_2 a_2 a_0 e^{i \omega_0 \tau}, \quad T_-' = g_1 a_1 a_0^\dagger e^{- i \omega_0 \tau}.
\end{align}

Consequently, these channels directly yield 
\begin{align}
    \text{SB}_+ = g_2^2 \, U, \quad \text{SB}_- = g_1^2 \, V, 
\end{align}
such that the tightened certification threshold is given by
\begin{align}
    \abs{Q}^2  > U_{dc}  V_{dc} \geq UV,
\end{align}
where the spectral shaping approach leads to the sharper entanglement bound. This can be achieved by a spectral mask of individual harmonics using gratings or monochromators~\cite{norin2004design}.

\subsection{\label{app:GME_proof}GME certification proof}

In the following we prove the GME certification bound from Eq.~\eqref{eq:GME_bound}, which provides the biseparable bound  
\begin{align}
    \abs{Q} \leq H(U,V),
\end{align}
where we define $c = 2-\sqrt{3}$, and
\begin{align}
    H(U,V) & = \sqrt{cV} + (1+\sqrt{2}) \sqrt{U(V+1)} + \sqrt{2 UV}.
\end{align}

This bound holds for every biseparable three-mode state with finite relevant moments regardless of the Fock space support.

\emph{Proof.} To bound the expectation value $Q = \expval*{a_2^\dagger a_1 a_0^2}$, we identify that the measured operator acts on the Fock state expansion, with mode ordering $(012)$, as
\begin{align}
    a_2^\dagger a_1 a_0^2 \ket{t,s,r} = \sqrt{st(r+1)(t-1)} \ket{t-2, s-1,r+1}.
\end{align}

Hence, $Q$ does contain many different coherences, and we separate the contributions in order to bound the total expectation value. We first select those with $t=2$, $r=0$, such that $\ket{2,s,0} \to \ket{0,s-1,1}$, with the total contribution of these coherences being 
\begin{align}
    q_L =\sqrt{2} \expval{a_1 \otimes M_{02}},
\end{align}
with the measurement operator isolating the particular coherence $M_{02} =\dyad{0,1}{2,0}_{02}$, acting on the modes $(02)$. 
Note that the selected coherence $q_L$ is evaluated in the full optical state, and no restrictions of its support are assumed. Particularly, all remaining contributions to the coherence are bounded by terms dependending on $U$.  
We can now decompose the total expectation value in terms of the contributions
\begin{align}
    Q = q_L + R_A + R_B,
\end{align}
where $q_L$ accounts for the occupations $t=2$, $r=0$, while $R_A$ and $R_B$ consider the remaining contributions of $t \geq 3$, $r \geq 0$, and $t =2$, $r\geq 1$, respectively. Note that we classify the contributions by $t$ and $r$, while leaving the lower XUV occupation $s$ unrestricted. This accounts for all terms contributing to $Q$ since any coherence must annihilate two IR photons, and therefore starts at $t \geq 2$.
To bound the individual contributions we use $V = \expval{N_1(N_0 + 1)} =  n + d$, where we have defined $n = \expval{N_1}$ and $d = \expval{N_1 N_0}$.
Now, for every state, positivity of the density matrix and Cauchy-Schwarz gives
\begin{align}
    \abs{R_A} \leq \sqrt{2d U}, \quad \abs{R_B} \leq \sqrt{2 U (n + 1)},
\end{align}
and consequently
\begin{align}
    \abs{Q-q_L} \leq \sqrt{2dU}+ \sqrt{2U (n + 1)}.
\end{align}

We have used that positivity of the density matrix bounds the coherences by geometric means using Cauchy-Schwarz.
Finally, we bound the expression for biseparable states, which enters via $|q_L|$.
First consider product states across each of the three cuts. All expectations in the following estimates are evaluated in the product state under consideration.

\paragraph{Cut $(1|02)$.} For $\rho = \rho_1 \otimes \rho_{02}$, define $z=\bra{2,0}\rho_{02}\ket{2,0}$ and $w = \bra{0,1}\rho_{02}\ket{0,1}$. Positivity and $|\expval*{ a_1}|^2 \leq n$ give $|q_L|^2 \leq 2 n z w$, and $z + w \leq 1$. Factorization also gives $V = n (1+ \expval*{N_0}) \geq n (1 + 2 z)$. For $V>0$,
\begin{equation}
    \frac{|q_L|^2}{V} \leq \frac{2 z (1-z)}{1 + 2 z} \leq c, \quad c = 2 - \sqrt{3}.
\end{equation}

The maximum is attained at $z = (\sqrt{3} - 1)/2$. If $V=0$, then $n = 0$ and $q_L = 0$, so $|q_L|^2 \leq cV$ also holds.

\paragraph{Cut $(2|01)$.} For $\rho = \rho_2 \otimes \rho_{01}$,
\begin{equation}
    q_L = \sqrt{2} \, \expval{ |1\rangle\langle0|_2 } \expval{ a_1 \otimes |0 \rangle \langle 2|_0}.
\end{equation}

The first factor satisfies $|\langle |1\rangle\langle 0|_2\rangle|^2 \leq \expval*{N_2}$. For the second factor, apply Cauchy-Schwarz with $A = a_1^\dagger \otimes  |0\rangle\langle0|_0$, and $B=|0\rangle\langle2|_0$, such that
\begin{align}
    |\langle a_1 \otimes |0\rangle\langle2|_0 \rangle|^2 \le(n + 1)\langle N_0\rangle/2.
\end{align}

Since $U = \expval*{N_2} \expval{N_0}$ on this cut $|q_L|^2 \leq U(n+1)$.

\paragraph{Cut $(0|12)$.}
For $\rho=\rho_0\otimes\rho_{12}$,
\begin{equation}
    q_L = \sqrt{2} \, \langle |0 \rangle\langle 2|_0 \rangle \langle a_1 \otimes | 1 \rangle\langle 0|_2 \rangle.
\end{equation}

Positivity gives $| \langle |0 \rangle\langle 2|_0 \rangle|^2 \leq \langle N_0 \rangle /2$. Cauchy-Schwarz with $A = |0 \rangle\langle 1|_2$, $B = a_1$ gives
\begin{equation}
    |\langle a_1 \otimes |1 \rangle\langle 0|_2 \rangle|^2 \leq \langle |1 \rangle\langle 1|_2 \rangle \, n \leq \expval*{N_2} n.
\end{equation}

Factorization of $U$ on this cut therefore yields
\begin{equation}
    |q_L|^2 \leq U n \leq U (n + 1).
\end{equation}

\paragraph{Mixtures of different cuts.}
Consider a biseparable state $\rho = \sum_\lambda p_\lambda \rho_\lambda$, where each normalized $\rho_\lambda$ is separable across one of the cuts. Let $A$ collect components of cut $(1|02)$ and let $B$ collect those of the other two cuts. Write $w_A = \sum_{\lambda \in A} p_\lambda$ and $w_B = \sum_{\lambda \in B} p_\lambda$. Linearity, the triangle inequality, and weighted Cauchy-Schwarz give
\begin{equation}
\begin{aligned}
    |q_L| & \leq \sum_{\lambda\in A} p_\lambda \sqrt{cV_\lambda} + \sum_{\lambda \in B}p_\lambda \sqrt{U_\lambda (n_\lambda + 1)} \\
    & \leq \sqrt{cV} + \sqrt{U(n + 1)}.
\end{aligned}
\end{equation}

All partial population sums are nonnegative and bounded by the total moments, while $w_A , w_B \leq 1$. Thus the estimate applies to mixtures of different bipartitions, not only to states separable across a fixed cut. Together with the bounds on $R_A$ and $R_B$, we find
\begin{equation}
    |Q| \leq \sqrt{cV} + (1 + \sqrt{2} ) \sqrt{U (n + 1)} + \sqrt{2Ud},
\end{equation}
and using that $n,d \leq V$ yields
\begin{align}
    \abs{Q} \leq H(U,V),
\end{align}
which concludes the proof.

\subsection{\label{app:XX_proof}Reduced XUV-XUV entanglement}

Under the assumption of an independently prepared IR field state $\rho_0$ we have the total state $\rho = \rho_0 \otimes \rho_{12}$, with the XUV field state of the two harmonic modes $\rho_{12}$. For an IR field with non-zero coherence we set $C_0 = \expval*{a_0^2}$, and the measured RABBIT coherence factorizes $Q = C_0 C_{21}$, where $C_{21} = \expval*{a_2^\dagger a_1}$.
To find a witness for XUV-XUV entanglement we need, in addition to the RABBIT coherence $C_{21}$, the cross-population $J_{21} = \expval*{N_2N_1}$.
With this, the criteria for XUV-XUV entanglement is given by
\begin{align}
    \abs{C_{21}}^2 > J_{21} .
\end{align}

\emph{Proof.} For a separable two-mode harmonic state 
\begin{align}
    \rho_{12} = \sum_\lambda p_\lambda \rho_1^{(\lambda)} \otimes \rho_2^{(\lambda)},
\end{align}
we can write $C_{21} = \sum_\lambda p_\lambda \expval*{a_2^\dagger }_\lambda \expval*{a_1}_\lambda$, and convexity with $|\expval*{a_j}|^2 \leq \expval*{N_j}$ gives
\begin{equation}
\begin{aligned}
    \abs{C_{21}}^2 & \leq \sum_\lambda p_\lambda \abs{\expval*{a_2}_\lambda}^2 \abs{\expval*{a_1}_\lambda}^2 \\
    & \leq \sum_\lambda p_\lambda \expval*{N_2}_\lambda \expval*{N_1}_\lambda \\
    & = \expval{N_2 N_1}.
\end{aligned}
\end{equation}

Violation therefore excludes separability. Note that the assumption of the independent IR reference field is necessary for extracting $C_{21}$ from $Q$. 
However, the XUV cross-population $J_{21}$ requires additional measurements such as electron coincidence detection or auxiliary optical measurements.

\end{document}